\newif\ifelsstyle
\IfFileExists{elsarticle.cls}{%
  \elsstyletrue\documentclass[final,5p,times,twocolumn,numbers,sort&compress]{elsarticle}
}{%
  \elsstylefalse\documentclass[10pt,a4paper,twocolumn]{article}
}
\usepackage[T1]{fontenc}
\usepackage{amsmath,amssymb,graphicx,booktabs,array,microtype,xcolor}
\usepackage{mathptmx}

\ifelsstyle
  \journal{Physics Letters B}
\else
  \usepackage[top=19mm,bottom=21mm,left=14mm,right=14mm,columnsep=6mm]{geometry}
  \usepackage[numbers,sort&compress]{natbib}
\fi

\usepackage[colorlinks=true,linkcolor=blue,citecolor=blue,urlcolor=blue]{hyperref}
\newcommand{\MeV}{\mathrm{MeV}}
\newcommand{\keV}{\mathrm{keV}}
\newcommand{\ER}{E_R}
\newcommand{\Est}{E_\star}
\newcommand{\tU}{t_U}
\newcommand{\dd}{\mathrm{d}}
\newcommand{\papertitle}{Decaying Dark Matter for the Exothermic Recoil at LZ}
\newcommand{\paperabstract}{%
The LUX-ZEPLIN (LZ) experiment has reported a single nuclear-recoil candidate
at $248\pm23\,(\mathrm{stat})\pm23\,(\mathrm{sys})$ keV. We suggest that it
comes from a GeV-scale dark state $\chi_1$ that downscatters exothermically,
$\chi_1A\to\chi_2A$, and is continuously replenished in the halo by the slow
decay $\chi_0\to\chi_1\chi_1$ of a long-lived parent. Because the recoil energy
is supplied by the mass splitting rather than by the halo velocity, the event
is a narrow line with no low-energy tail. Kinematics requires
$m_{\chi_1}\gtrsim247$ MeV and fixes the splitting for each $m_{\chi_1}$. The
same line should appear at 700--810 keV in argon for
$m_{\chi_1}=0.3$--10 GeV. Normalizing to one LZ event also fixes the local
rate of the radiative decay $\chi_1\to\chi_2\gamma\gamma$, up to the
mediator's photon coupling. A rough $\gamma$-ray estimate then requires this
coupling to be $\lesssim10^{-2}$ of a proton-loop reference at
$m_{\chi_1}=10$ GeV, but $\lesssim10^{-9}$ at 0.3 GeV, which favours daughters
near 10 GeV.}

\begin{document}
\ifelsstyle
\begin{frontmatter}
\title{\papertitle}
\author[1]{Yongsoo Jho}
\ead{1jys34@gmail.com}
\author[1]{Sanghwan Kim}
\ead{sanghwankim97@yonsei.ac.kr}
\author[1,2]{Seong Chan Park}
\ead{sc.park@yonsei.ac.kr}
\affiliation[1]{organization={Department of Physics and IPAP, Yonsei University},
  city={Seoul}, postcode={03722}, country={South Korea}}
\affiliation[2]{organization={School of Physics, Korea Institute for Advanced Study},
  addressline={85 Hoegi-ro}, city={Seoul}, postcode={02455}, country={South Korea}}
\begin{abstract}\paperabstract\end{abstract}
\end{frontmatter}
\else
\twocolumn[{
\begin{center}
{\LARGE\bfseries\papertitle\par}\vspace{8pt}
{\normalsize Seong Chan Park}\par
{\small Department of Physics and IPAP, Yonsei University, Seoul 03722, South Korea;
School of Physics, KIAS, Seoul 02455, South Korea}\par\vspace{8pt}
\end{center}
\noindent\textbf{Abstract}\par\smallskip
\noindent\paperabstract\par\medskip
\par\vspace{10pt}\hrule\vspace{12pt}
}]
\fi

\section{Introduction}
\label{sec:intro}
With 2.84 tonne-years of exposure, LUX-ZEPLIN (LZ) reports one candidate at
$\ER=248\pm23\,(\mathrm{stat})\pm23\,(\mathrm{sys})\,\keV$ in its extended
nuclear-recoil search. Across the models tested by LZ, the background-only
hypothesis shows a global tension of $2.6\sigma$~\cite{LZ2026}. One event does
not establish a signal, but its high energy and the absence of a corresponding
low-energy excess motivate mechanisms with a hard or narrow spectrum. The
candidate has prompted a large number of interpretations. These include endothermic and
exothermic inelastic DM~\cite{SmithWeiner,Graham,Barello}, often realized
with Higgsinos, other electroweak multiplets or dark-photon and scalar
portals~\cite{DiMauro2026,deLima2026,BaerBarger2026,LZth260926570,LZth260923691,LZth260922739,LZth260922063,LZth260921011,LZth260918564,LZth260917196,LZth260917412,LZth260915027,LZth260915413,LZth260915600,LZth260915742,LZth260915714,LZth260913038,LZth260910491,LZth260910453,LZth260910827,LZth260908893,LZth260908993,LZth260909015,LZth260909138,LZth260909385,LZth260906909,LZth260907138,LZth260907225,LZth260907451,LZth260907800,LZth260907811,LZth260906571,LZth260906825,LZth260906171,LZth260904144,LZth260904186,LZth260904163,LZth260902807,LZth260902868,LZth260901475,LZth260901583,LZth260901590,LZth260901504,LZth260901892}; boosted
or high-velocity DM and neutrino-induced
recoils~\cite{Kannike2026,Liang2026,Agashe,Jho2020,Jho2021,LZth260924982,LZth260921444,LZth260911600,LZth260910504,LZth260906890,LZth260904185};
composite, extra-dimensional and other non-standard
mechanisms~\cite{Jung2026,LZth260924988,LZth260923477,LZth260925114,LZth260923096,LZth260915118,LZth260915933,LZth260912045,LZth260909037,LZth260909107,LZth260909136}; and nuclear-response
effects~\cite{Khan2026}. Their implications for solar capture, sideband and
multi-target data, halo modelling and colliders have been examined in
Refs.~\cite{Mahapatra2026,LZth260926698,LZth260921823,LZth260916529,LZth260915321,LZth260915634,LZth260915985,LZth260919174,LZth260911833,LZth260909830,LZth260910636,LZth260908712,LZth260907807,LZth260906640,LZth260906760,LZth260906750,LZth260904673,LZth260905291,LZth260904181,LZth260904175,LZth260902775}. Galactic production of excited
states by endothermic upscattering was considered in
Ref.~\cite{LZth260917935}, a mechanism distinct from the decay regeneration
proposed here. In contrast to heavy exothermic
DM~\cite{deLima2026,BaerBarger2026}, our $\chi_1$ is GeV-scale and $\delta$
is MeV-scale, which changes the target dependence of the line
(Table~\ref{tab:exo}); in contrast to
boosted scenarios~\cite{Kannike2026,Liang2026}, the recoil energy is supplied
by the mass splitting rather than by the incident momentum.

Exothermic scattering $\chi_1A\to\chi_2A$ with $\delta=m_{\chi_1}-m_{\chi_2}>0$ turns the
internal energy into a recoil line whose position is nearly independent of
the halo velocity. For heavy DM this requires $\delta\simeq\ER$, as in
Refs.~\cite{deLima2026,BaerBarger2026}. For a GeV-scale $\chi_1$, momentum
conservation instead requires $\delta$ of MeV size or larger. A primordial excited
population with such a splitting may be depleted by decays or by
$\chi_1\chi_1\to\chi_2\chi_2$ conversion~\cite{Graham,BatellPospelovRitz},
depending on the couplings and the thermal history. Here we observe that a
late decay
\begin{equation}
 \chi_0\to\chi_1\chi_1,\qquad \chi_1+A\to\chi_2+A,
 \label{eq:process}
\end{equation}
of a long-lived parent replenishes the excited population today. We assume
that the primordial $\chi_1$ abundance is negligible, so that the present
population is dominated by parent decays. Decays of
long-lived DM have also been invoked for, and constrained by, other
anomalous signals~\cite{Jho2025KM3,Rott2015}. A small
release $\epsilon=m_{\chi_0}-2m_{\chi_1}$ keeps most of the daughters bound to the halo, so
the recoil line stays narrow.

We derive the kinematic conditions, test the sensitivity of the line to the
xenon nuclear response, and show that the coupling needed for one LZ event
fixes the local emissivity of $\chi_1$ radiative decays. Illustrative photon
sensitivities then differ greatly between sub-GeV and 10 GeV daughters.

\section{Kinematics}
\label{sec:kin}
\begin{figure*}[t]
\centering\includegraphics[width=0.92\textwidth]{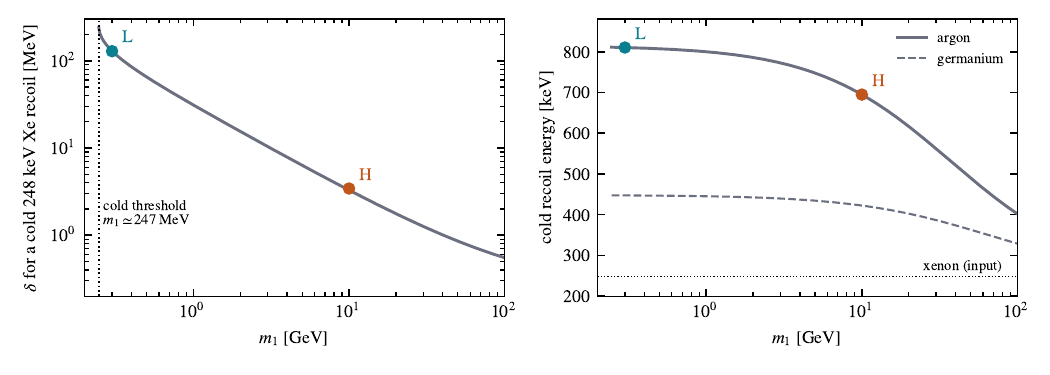}
\caption{Left: splitting $\delta_0$ that yields a cold 248 keV recoil on
natural xenon, eq.~\eqref{eq:threshold}; below the dotted threshold no
$m_{\chi_2}\geq0$ exists. Right: the corresponding cold line in argon and germanium,
eq.~\eqref{eq:target}, for a cold xenon line fixed at 248 keV. Dots mark the
masses of benchmarks L and H; Table~\ref{tab:bench} gives the values for their
optimized parameters.}
\label{fig:kin}
\end{figure*}

We write $m_{\chi_0}=2m_{\chi_1}+\epsilon$ and $m_{\chi_2}=m_{\chi_1}-\delta$. For a parent at rest,
each daughter has kinetic energy $\epsilon/2$ and speed
$v_{\rm kick}\simeq\sqrt{\epsilon/m_{\chi_1}}$. For an incident $\chi_1$ with energy
$E_1$ and momentum $p_1$ on a nucleus of mass $M_A$ at rest, the recoil energy
is exactly
\begin{equation}
 \ER=C+H\cos\theta^*,\quad H=\frac{p_1p_f}{\sqrt s},\quad
 C=\gamma_{\rm cm}\sqrt{M_A^2+p_f^2}-M_A,
 \label{eq:exact}
\end{equation}
where $s=m_{\chi_1}^2+M_A^2+2M_AE_1$, $\gamma_{\rm cm}=(E_1+M_A)/\sqrt s$,
$p_f=\lambda^{1/2}(s,m_{\chi_2}^2,M_A^2)/(2\sqrt s)$, and $\theta^*$ is the
centre-of-mass recoil angle. For $p_1\to0$ the band collapses to the cold line
\begin{equation}
 E_A^{(0)}=\frac{m_{\chi_1}^2-m_{\chi_2}^2}{2(M_A+m_{\chi_1})}
 =\frac{\delta(2m_{\chi_1}-\delta)}{2(M_A+m_{\chi_1})}.
 \label{eq:cold}
\end{equation}
Requiring $E_A^{(0)}=\Est$ with $m_{\chi_2}\geq0$ gives
\begin{equation}
 m_{\chi_1}\geq \Est+\sqrt{\Est^2+2M_A\Est},\quad
 \delta_0=m_{\chi_1}-\sqrt{m_{\chi_1}^2-2(M_A+m_{\chi_1})\Est}.
 \label{eq:threshold}
\end{equation}
For natural xenon ($\bar M_A=122.3$ GeV) and $\Est=248$ keV, this
cold-limit threshold is $m_{\chi_1}\simeq246.5$ MeV. It ranges from 243 to 251 MeV across the isotopes with more than 1\% abundance.
The momentum transfer is $q\simeq\sqrt{2M_A\Est}\simeq246$ MeV. The left panel
of Fig.~\ref{fig:kin} shows $\delta_0(m_{\chi_1})$: $\delta_0/m_{\chi_1}$ falls from 0.43 at
0.3 GeV to $3.3\times10^{-4}$ at 10 GeV.

Lighter states are not kinematically forbidden if they are boosted, but then
the incident momentum has to supply the recoil, so that
$H/C\gtrsim[1-m_{\chi_1}^4/(4M_A^2C^2)]^{1/2}\to1$ and the recoil band extends to 
zero energy. For $m_{\chi_1}=100$ MeV and a band centred on 248 keV, for example, $H/C>0.98$,
and with the Helm form factor and the contact interaction adopted below, 98\%
of the events fall below 100 keV (without nuclear suppression the recoil band
would be flat). Above the
threshold, a small $p_1$ gives a narrow line. For a velocity $v_1$ the
half-width is $H\simeq\mu_Av_1\sqrt{2\Est/M_A}$, i.e.\ about 19 keV at
$m_{\chi_1}=10$ GeV and $v_1=300$ km/s, and it is much smaller for $m_{\chi_1}\lesssim1$ GeV.
A large $\epsilon$ only broadens the line.\footnote{For $m_{\chi_1}=10$ GeV and
$\epsilon=100$ keV the standard deviation of the true spectrum grows from 11.5
to 41 keV, and 17.5\% of the events move above 270 keV.}

Eq.~\eqref{eq:cold} also gives a target-independent prediction,
\begin{equation}
 \frac{E_B^{(0)}}{E_A^{(0)}}=\frac{M_A+m_{\chi_1}}{M_B+m_{\chi_1}}.
 \label{eq:target}
\end{equation}
For a cold xenon line fixed at 248 keV and GeV-scale $m_{\chi_1}$, this gives an
argon line at 810 keV ($m_{\chi_1}=0.3$ GeV) or 695 keV ($m_{\chi_1}=10$ GeV), and a
germanium line at 420--450 keV (Fig.~\ref{fig:kin}, right). With the optimized
parameters of Table~\ref{tab:bench} the cold argon lines are 808 and 726 keV;
form-factor weighting moves the mean of the argon spectrum for H to about
760 keV. Heavy exothermic DM instead gives
$E_{\rm Ar}/E_{\rm Xe}\to1$~\cite{BaerBarger2026,LZth260915782}, or intermediate values for
$m_{\chi_1}\sim30$--$200$ GeV~\cite{deLima2026}. A combined xenon--argon line measurement can therefore constrain $m_{\chi_1}$, once
form-factor weighting and velocity broadening are taken into account.

\section{Comparison with other interpretations}
\label{sec:compare}
Either sign of the splitting can give a high-energy recoil, but the two
cases are powered differently. In endothermic up-scattering
$\chi_lA\to\chi_hA$ of a halo state, the energy comes from the incident
motion. This requires $\mu_Av^2/2>\delta$, where $\mu_A$ is a reduced mass, $\mu_A = (m_1m_A)/(m_1+m_A)$, and the largest recoil is 
\begin{equation}
 E_R^{+}=\frac{\mu_A^2v^2}{2M_A}\Big[1+\Big(1-\frac{2\delta}{\mu_Av^2}\Big)^{1/2}\Big]^2 .
 \label{eq:endo}
\end{equation}
With $v\leq v_{\max}\simeq776$ km/s, eq.~\eqref{eq:endo} reaches 248 keV only if
$\mu_A\gtrsim48$ GeV, i.e.\ $m_\chi\gtrsim80$ GeV even for $\delta\to0$; a
10 GeV state gives at most 9 keV. For TeV-scale DM with
$\delta\simeq0.3$ MeV~\cite{DiMauro2026} only $v\gtrsim700$ km/s contributes,
and the recoils spread over roughly 100--750 keV. The rate is then controlled by
the high-velocity tail of the halo~\cite{LZth260921444}.

Alhazmi et al.~\cite{LZth260906890} avoid this dependence with endothermic
scattering of \emph{boosted} light DM. A nearly monochromatic flux produced in
the dark sector scatters close to threshold, so the recoils are confined
between two kinematic endpoints, and the up-scattered state may decay
invisibly. In the notation of eq.~\eqref{eq:exact} this is the limit
$p_f\to0$, where $H\to0$ and $\ER\to M_A(\gamma_{\rm cm}-1)$. The line position
is then set by the incident momentum, which must be relativistic
($p_1\simeq q\simeq246$ MeV), and its width by the distance above threshold.
A narrow line therefore requires a monochromatic flux tuned close to
threshold. Because the threshold depends on $M_A$, the same flux is generally
not near threshold in argon.

In exothermic downscattering the energy comes from the mass splitting instead.
The recoil is the line of eq.~\eqref{eq:cold}: it needs neither a velocity tail
nor a boost, its position is fixed by $\delta$ rather than by the incident
momentum, and its width is set only by the non-relativistic motion of $\chi_1$.
It is available for any $m_{\chi_1}$ above the cold threshold, including GeV-scale
states, and gives the target dependence of eq.~\eqref{eq:target}.

The price of the exothermic choice is that the halo must contain the excited
state, and the exothermic interpretations proposed so far differ mainly in
where that state comes from (Table~\ref{tab:exo}).
Refs.~\cite{BaerBarger2026,deLima2026} take a relic in the excited state of a
pseudo-Dirac pair: the fraction that survives
$\chi_h\chi_h\to\chi_l\chi_l$ conversion after freeze-out,
$f_h\simeq(2$--$40)\times10^{-3}$, or a freeze-in population at low reheating
temperature. Survival over a Hubble time then requires $\delta<2m_e$, which
closes $\chi_h\to\chi_le^+e^-$, and with $\delta\lesssim1$ MeV the line of
eq.~\eqref{eq:cold} reaches 248 keV only for $m_\chi\gtrsim40$ GeV.
Ref.~\cite{LZth260917935} instead regenerates the excited state in the
Galaxy by endothermic self-scattering $\chi_1\chi_1\to\chi_2\chi_2$ through
a light scalar, which needs a strong self-interaction and again TeV-scale
masses with sub-MeV splittings. Decay regeneration removes the longevity
requirement differently: the excited state is replenished continuously, so
$\delta$ may exceed $2m_e$ by orders of magnitude. This is what opens the
GeV mass range and gives the target ratio of eq.~\eqref{eq:target}; in
return, the same splitting drives $\chi_1\to\chi_2\gamma\gamma$, whose
local emissivity is fixed by the LZ rate (Sec.~\ref{sec:viability}). The
observable differences are the width of the xenon line and, most directly,
the argon energy: $E_{\rm Ar}/E_{\rm Xe}\simeq1.1$--1.3 for TeV-scale
states, 1.4--2.3 for Ref.~\cite{deLima2026}, and 2.8--3.3 here.

\begin{table*}[t]
\centering\small
\setlength{\tabcolsep}{4pt}
\begin{tabular}{>{\raggedright\arraybackslash}p{0.14\textwidth}>{\raggedright\arraybackslash}p{0.24\textwidth}>{\raggedright\arraybackslash}p{0.24\textwidth}>{\raggedright\arraybackslash}p{0.28\textwidth}}
\toprule
 & relic excited state, heavy~\cite{BaerBarger2026} & relic excited state, intermediate~\cite{deLima2026} & decay regeneration (this work)\\
\midrule
scattering state; $\delta$ & 0.3--1 TeV; 0.275--0.4 MeV & 30--200 GeV; 0.5--1 MeV & $m_{\chi_1}\gtrsim0.25$ GeV (benchmarks 0.3, 10 GeV); 3.4--130 MeV, fixed by $m_{\chi_1}$ [eq.~\eqref{eq:threshold}]\\
origin of the excited state & primordial: residue of $\chi_h\chi_h\to\chi_l\chi_l$ conversion after freeze-out & primordial: freeze-out residue, or freeze-in with $T_{\rm RH}\simeq m_\chi/9$ & continuous: late $\chi_0\to\chi_1\chi_1$ decay with small $\epsilon$\\
excited fraction today & $2\times10^{-3}$--$4\times10^{-2}$ & $6\times10^{-3}$ at the benchmark & free; 6.7\% in the benchmark of Sec.~\ref{sec:viability}\\
de-excitation & must survive $\tU$: $\delta<2m_e$; $\gamma\gamma$, $\nu\bar\nu$ slower than Hubble & must survive $\tU$: $\delta<2m_e$ & need not survive $\tU$; $\chi_1\to\chi_2\gamma\gamma$ emissivity fixed by the LZ rate, $r_\gamma\lesssim0.02$ at 10 GeV (Sec.~\ref{sec:viability})\\
xenon spectrum & halo-broadened; peaks near 160 keV, event on its upper side & halo-broadened; peak $\simeq270$ keV, support $\simeq140$--530 keV at the benchmark & narrow line, $\sigma_R\simeq4$--12 keV, no low-energy tail\\
$E_{\rm Ar}/E_{\rm Xe}$, eq.~\eqref{eq:target} & 1.1--1.3 (argon line at 268--311 keV in Ref.~\cite{LZth260915782}) & 1.4--2.3 (2.0 at 45 GeV) & 2.8--3.3\\
\bottomrule
\end{tabular}
\caption{Exothermic interpretations of the LZ candidate. The entries for
Refs.~\cite{BaerBarger2026,deLima2026} are taken from those papers; the
last row evaluates the cold-line ratio of eq.~\eqref{eq:target} over each
mass range and neglects form-factor weighting and velocity broadening.}
\label{tab:exo}
\end{table*}

The late decay~\eqref{eq:process} supplies the excited state with a small
kick, and its radiative decay must then be controlled
(Sec.~\ref{sec:viability}). A small decay kick cannot instead power
endothermic $\chi_2A\to\chi_1A$: in the non-relativistic limit the incident
$\chi_2$ needs at least $\delta(1+m_{\chi_2}/M_A)$ of kinetic energy. In the
present scenario the decay supplies the excited population, while its
splitting supplies the recoil energy.

\section{Interaction and nuclear response}
\label{sec:model}
As a concrete realization we take real scalars $\chi_{0,1,2}$ and a scalar
mediator $\phi$,
\begin{equation}
 \mathcal L\supset
 -\tfrac12 g_{011}\chi_0\chi_1^2-g_{12}\phi\chi_1\chi_2
 -\phi\,(g_p\bar pp+g_n\bar nn),
 \label{eq:model}
\end{equation}
with $\chi_{1,2}$ odd under a $Z_2$. We take $m_\phi^2\gg q^2\simeq(246\,\MeV)^2$,
which also closes $\chi_1\to\chi_2\phi$, and assume no tree-level coupling of
$\phi$ to electrons. For $g_p=g_n\equiv g_N$ and coherent
scattering, the rate per target nucleus is
\begin{equation}
 v_1\sigma_A=\sigma_n\frac{m_{\chi_1}^2}{\mu_n^2}
 \left\langle\frac{A^2M_Ap_f}{\sqrt s\,E_1}F_A^2(\ER)\right\rangle_{\theta^*},
 \quad
 \sigma_n\equiv\frac{g_{12}^2g_N^2\mu_n^2}{4\pi m_{\chi_1}^2m_\phi^4},
 \label{eq:vsigma}
\end{equation}
where $\mu_n$ is the $\chi_1$--nucleon reduced mass, $\sigma_n$ is the
corresponding elastic-limit nucleon cross section, and the average runs over
$\cos\theta^*$. In the cold limit eq.~\eqref{eq:vsigma} becomes
$v_1\sigma_A=\sigma_nA^2(\mu_A/\mu_n)^2(p_f/\mu_A)F_A^2$, the familiar
coherent form with the velocity replaced by $p_f/\mu_A$. This factor is 0.03
at $m_{\chi_1}=10$ GeV and 0.8 at 0.3 GeV.

We use the Helm form factor~\cite{LewinSmith} with natural isotope
abundances~\cite{NIST}. The candidate lies just below the second Helm zero,
which falls at 263, 276 and 287 keV for $^{136}$Xe, $^{132}$Xe and $^{129}$Xe
(Fig.~\ref{fig:nuc}, left), and there $F_A^2\sim10^{-4}$. The shape and rate
near 248 keV therefore depend on the nuclear model. Shell-model responses shift
and partly fill such minima~\cite{Anand}, and interference between operators
can move them substantially~\cite{Khan2026}. As sensitivity tests, not a calibrated nuclear uncertainty, we shift the
Helm parameter $c_A$ by $\pm0.2$ fm and $s$ by $\pm0.1$ fm.

\section{Spectral benchmarks}
\label{sec:spectra}
Parent velocities follow a truncated Maxwellian ($v_0=220$, $v_{\rm esc}=544$,
$v_E=232$ km/s). The daughters are boosted exactly and scattered with the weight
of eq.~\eqref{eq:vsigma}. As an energy-only figure of merit we use
\begin{equation}
 K=\int^{C + \sigma_R}_{C-\sigma_R} \dd\ER\,f(\ER)\,e^{-(\ER-\Est)^2/2\sigma_\star^2},\qquad
 \sigma_\star=32.5\ \keV,
 \label{eq:K}
\end{equation}
with $f$ the unit-normalized true spectrum, so that $K=1$ is an ideal line at
$\Est$. At fixed $(m_{\chi_1},\epsilon)$ we choose $\delta$ to maximize $K$. The
quoted uncertainties are added in quadrature. This procedure locates the line;
it does not replace the detector-level likelihood of Ref.~\cite{LZ2026}.

\begin{table}[t]
\centering\small
\setlength{\tabcolsep}{4pt}
\begin{tabular}{lcc}
\toprule
 & L & H \\
\midrule
$m_{\chi_1}$ [GeV] & 0.3 & 10 \\
$\epsilon$ [keV]; $v_{\rm kick}$ [km/s] & 0.1; 173 & 1; 95 \\
$\delta$ [MeV] & 128.6 & 3.43 \\
$\sigma_R$ [keV]; 90\% range [keV] & 3.8; 239--253 & 11.5; 230--266 \\
$K$ & 0.993 & 0.944 \\
mean shift at fixed $\delta$ [keV] & $<1$ & $-1$ to $+6$ \\
rate factor at fixed $\delta$ & 0.2--2.5 & 0.3--2.8 \\
shift of refitted $\delta$ & $<0.4\%$ & $<3\%$ \\
$\sigma_nf_1$ [cm$^2$] for 1 event & $4\times10^{-47}$ & $1.1\times10^{-45}$ \\
$T_1$ [yr] & 73 & $2.9\times10^{12}$ \\
$\langle E_{\gamma\gamma}\rangle$; $E_{\gamma,\max}$ [MeV] & 119; 101 & 3.4; 3.4 \\
proxy $r_\gamma$, $\tau_\gamma=10^{27}$ s & $2\times10^{-9}$ & $1.7\times10^{-2}$ \\
cold $E_{\rm Ar}$ [keV] & 808 & 726 \\
\bottomrule
\end{tabular}
\caption{Benchmarks. $\sigma_R$ and the range are for the true xenon spectrum.
Nuclear variations are the Helm sensitivity tests of Sec.~\ref{sec:model}.
The event rate assumes $\rho_\odot=0.3$ GeV\,cm$^{-3}$ and an illustrative
50\% acceptance. $\langle E_{\gamma\gamma}\rangle$ is the mean total energy
carried by both photons and $E_{\gamma,\max}=(m_{\chi_1}^2-m_{\chi_2}^2)/2m_{\chi_1}$ the single-photon endpoint.
$T_1$ is the $\chi_1\to\chi_2\gamma\gamma$ lifetime at
$\sigma_n=\sigma_nf_1|_{\rm 1\,event}$ with the proton-loop photon coupling;
$r_\gamma$ is defined in Sec.~\ref{sec:viability}; its tabulated value uses a
two-photon line limit as an illustrative proxy for a continuum signal.}
\label{tab:bench}
\end{table}

\begin{figure*}[t]
\centering\includegraphics[width=0.92\textwidth]{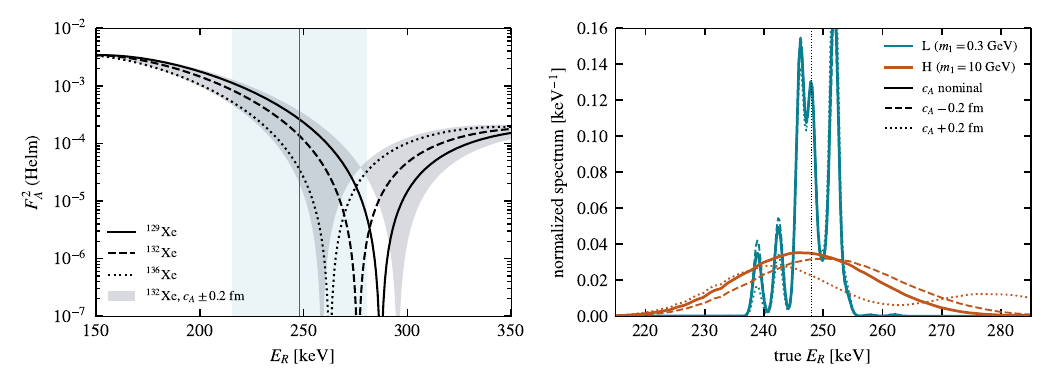}
\caption{Left: Helm $F_A^2$ for three xenon isotopes near the candidate
(vertical line; the shaded band is $\pm\sigma_\star$). The grey band shows
$c_A\pm0.2$ fm for $^{132}$Xe. Right: true recoil spectra of benchmarks L and
H for nominal and shifted $c_A$ at fixed particle parameters; a 0.5 keV
smoothing is applied for display.}
\label{fig:nuc}
\end{figure*}

Table~\ref{tab:bench} lists a light (L) and a heavy (H) benchmark.
Both kicks lie below the local escape speed; including the parent velocity,
2.5\% (L) and 0.7\% (H) of the daughters are unbound at injection. We assume
that the accumulated daughters keep the injected phase-space distribution
(parent halo convolved with the kick); orbital evolution could modify
it, mainly affecting the width for H.\footnote{A kick above the
local escape speed leaves an unbound population whose local density is
suppressed by the ratio of the halo crossing time to $\tau_0$. With
$\epsilon=1$ keV, benchmark L has $v_{\rm kick}\simeq550$ km/s and 62\% of
the daughters are unbound.}
For L, the isotope-dependent cold lines form a comb of total width
$\sigma_R\simeq4$ keV. For H, halo motion gives $\sigma_R\simeq12$ keV.
The optimal $\delta$ for H is 4.5\% above $\delta_0$, because $F_A^2$ falls
steeply towards the zero. We distinguish two effects of the nuclear variations. At fixed particle
parameters (Fig.~\ref{fig:nuc}, right), the mean recoil energy moves by less
than 1 keV for L and by up to 6 keV for H, whose shape also changes as the zero
moves across the band. If instead $\delta$ is refitted, it changes by less than
0.4\% (L) and 3\% (H). In both cases the rate changes by roughly an order of
magnitude. The line position is thus much less sensitive to the nuclear
response than the normalization, which inherits the full nuclear
uncertainty. After smearing with $\sigma_\star$ the two
benchmarks, and masses up to $\sim10$ GeV, are not meaningfully
distinguished by this energy-only comparison.\footnote{A mediator coupled only to neutrons changes the
optimal $\delta$ by $<0.05\%$ and multiplies the rate by
$(N/A)^2\simeq0.34$.}

\section{Rate, radiative decay, and an explicit benchmark}
\label{sec:viability}
Bound daughters accumulate. We measure all abundances as fractions of the
present local DM density $\rho_\odot$, neglect transport and the factor
$2m_{\chi_1}/m_{\chi_0}\simeq1$, and assume that the primordial $\chi_1$ abundance is
negligible. A parent with initial fraction $f_{0,i}$ then evolves as
$f_0(t)=f_{0,i}e^{-t/\tau_0}$, and
\begin{equation}
 f_1(\tU)=f_{0,i}\,\frac{\tau_1}{\tau_0-\tau_1}
 \left(e^{-\tU/\tau_0}-e^{-\tU/\tau_1}\right).
 \label{eq:f1}
\end{equation}
Decays that impart velocity kicks are constrained by Milky-Way
satellites~\cite{Mau2022}; that analysis concerns a massive daughter plus dark
radiation, whereas here the fractional mass loss $\epsilon/m_{\chi_0}\lesssim2\times10^{-7}$
is negligible and only the kick matters. We therefore treat $f_{0,i}$ and
$\tau_0$ as free parameters. One accepted event in 2.84 tonne-years then fixes
$\sigma_nf_1$ (Table~\ref{tab:bench}); this product scales inversely with the
acceptance and with the nuclear factor.

The same product $g_{12}g_N/m_\phi^2$ enters the radiative decay
$\chi_1\to\chi_2\gamma\gamma$ through an off-shell $\phi$ (the single-photon
transition is forbidden for spin-0 $\chi_{1,2}$); radiatively decaying scalar
dark matter has also been studied in other portal settings~\cite{Park2013}.
Hadronic channels are closed, since $\delta<2m_\pi$ throughout and a single
$\pi^0$ is forbidden by parity for a CP-even $\phi$. The nucleon-level
Lagrangian~\eqref{eq:model} does not fix the coefficient of
$\phi F_{\mu\nu}F^{\mu\nu}$, which depends on the quark-level couplings of
$\phi$.\footnote{\label{fn:rgamma}For a CP-even scalar coupled to quarks, the ratio of its
photon coupling to $g_N$ is set by $\sum_qN_cQ_q^2y_q/m_q$ against the
nucleon matrix elements, giving $r_\gamma\approx5$--20 for heavy-quark,
Higgs-like or light-quark couplings, so $r_\gamma\ll1$ requires a
cancellation among flavours. A vector mediator coupled to baryon
number removes $\chi_1\to\chi_2\gamma\gamma$ by C invariance but opens
$\chi_1\to\chi_2e^+e^-$ through kinetic mixing for $\delta>2m_e$.}
We parameterize it as $\mathcal L_{\rm eff}\supset-c_\gamma\phi
F_{\mu\nu}F^{\mu\nu}/4$, with $c_\gamma=2\alpha g_pr_\gamma/(3\pi m_p)$. Here
$r_\gamma=1$ denotes a point-like heavy-proton loop reference, not a
prediction of a UV model. It gives
$\Gamma_{\phi\to\gamma\gamma}(\sqrt s)=r_\gamma^2\alpha^2g_p^2s^{3/2}/
(144\pi^3m_p^2)$ and
\begin{equation}
 \Gamma_1=\frac{r_\gamma^2\,g_{12}^2g_p^2}{m_\phi^4}\int_0^{\delta^2}
 \frac{\dd s}{\pi}\,
 \frac{\lambda^{1/2}(m_{\chi_1}^2,m_{\chi_2}^2,s)}{16\pi m_{\chi_1}^3}\,
 \frac{\alpha^2s^2}{144\pi^3m_p^2}.
 \label{eq:gamma1}
\end{equation}
Let $T_1$ denote the $\chi_1$ lifetime for $r_\gamma=1$ at the cross section
$\sigma_n=\sigma_nf_1|_{\rm 1\,event}$. Since $\tau_1\propto1/\sigma_n\propto
f_1$, we have $\tau_1=(T_1/r_\gamma^2)f_1$. Inserting this into
eq.~\eqref{eq:f1}, the right-hand side becomes $f_{0,i}(T_1/r_\gamma^2)$ times
the slope of the chord of $e^{-\tU/y}$ between $y=\tau_1$ and $y=\tau_0$.
A positive solution therefore exists only if
\begin{equation}
 \frac{T_1}{r_\gamma^2}\geq\frac{B_{\min}(\tau_0)}{f_{0,i}},\qquad
 B_{\min}^{-1}=\max_{y>0}\frac{e^{-\tU/\tau_0}-e^{-\tU/y}}{\tau_0-y}.
 \label{eq:regen}
\end{equation}
Numerically, $B_{\min}=2.09\,\tU$ for $\tau_0=\tU$ and $10.9\,\tU$ for
$\tau_0=10\,\tU$, and $B_{\min}\to\tau_0e^{\tU/\tau_0}\simeq\tau_0+\tU$ for
$\tau_0\gg\tU$.
A continuously replenished $\chi_1$ need not itself survive for a Hubble time.

The same relation gives a local, production-independent result. At the Solar
position, the $\chi_1\to\chi_2\gamma\gamma$ decay rate per unit volume is
\begin{equation}
 \frac{n_1}{\tau_1}=\frac{\rho_\odot}{m_{\chi_1}}\frac{f_1}{\tau_1}
 =\frac{\rho_\odot}{m_{\chi_1}}\frac{r_\gamma^2}{T_1}.
 \label{eq:emissivity}
\end{equation}
The one-event scattering normalization fixes this local emissivity up to
$r_\gamma$ independently of $f_{0,i}$, $\tau_0$ and, more generally, the
production history of $\chi_1$. Since $n_1/\tau_1\propto[\sigma_nf_1]_{\rm LZ}\,
(c_\gamma/g_N)^2$, what is constrained is the ratio $c_\gamma/g_N$, which
$r_\gamma$ measures. Converting it to a Galactic flux requires the
spatial distribution of $\chi_1$; we assume it traces the halo for the
following estimate. As a rough sensitivity proxy, we compare the number of
decays with a two-photon line limit at a hypothetical decaying-DM mass equal
to the mean total photon energy $\langle E_{\gamma\gamma}\rangle$~\cite{Essig2013}.
Other Galactic photon signals from dark sectors have been considered in
Refs.~\cite{Huh2008,Jho2025halo}. The resulting estimate is
\begin{equation}
 r_\gamma^2\lesssim\min\left[\frac{f_{0,i}T_1}{B_{\min}},\
 \frac{T_1m_{\chi_1}}{\tau_\gamma\langle E_{\gamma\gamma}\rangle}\right].
 \label{eq:rgamma}
\end{equation}
Published lifetime limits span $\tau_\gamma\sim10^{24}$--$10^{28}$ s,
depending on mass and channel~\cite{Essig2013}. Our photons instead form a
three-body continuum; for benchmark L the single-photon endpoint is 101 MeV.
Matching the total emitted energy does not match the line spectrum or the
instrument response. Equation~\eqref{eq:rgamma} therefore supplies a scale
estimate, not an exclusion; a spectral recast is required.

\begin{figure}[t]
\centering\includegraphics[width=\columnwidth]{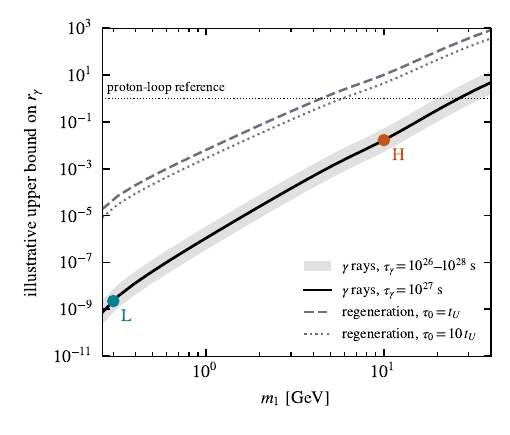}
\caption{Reference sensitivity estimates for the effective photon coupling
$r_\gamma$ of the mediator, relative to the proton-loop reference, from
eq.~\eqref{eq:rgamma}: MeV $\gamma$ rays for $\tau_\gamma=10^{27}$ s (solid;
band $10^{26}$--$10^{28}$ s), and the regeneration condition~\eqref{eq:regen}
for $f_{0,i}=1$ with $\tau_0=\tU$ (dashed) and $10\,\tU$ (dotted). At each $m_{\chi_1}$, $\delta$
is optimized and $\sigma_nf_1$ is fixed by one LZ event with 50\% acceptance.
The $\gamma$-ray curves use a two-photon line sensitivity as a proxy for the
actual three-body spectrum; they are not exclusion limits.}
\label{fig:photon}
\end{figure}

Figure~\ref{fig:photon} shows these sensitivity estimates. Because $\Gamma_1$
grows steeply with $\delta$, the photon proxy gives
$r_\gamma\lesssim2\times10^{-9}$ for benchmark L and $r_\gamma\sim10^{-6}$ at
1 GeV, but only $r_\gamma\lesssim0.02$ (0.005--0.05 across the band) at 10 GeV.
For the parent parameters displayed, the regeneration
condition~\eqref{eq:regen} is weaker throughout; it need not remain so for
arbitrarily large $B_{\min}/f_{0,i}$. The value $r_\gamma\simeq1$ crosses the
illustrative photon curve above $\sim25$ GeV, where halo motion broadens the
line ($K\simeq0.7$), although this energy-only measure does not exclude such
masses. For the unsuppressed values $r_\gamma\approx5$--20 of
footnote~\ref{fn:rgamma} the same proxy requires $m_{\chi_1}\gtrsim40$--60 GeV, where
$K\approx0.4$--0.5 and the cold argon line falls to 520--570 keV. A narrow
line therefore implies a suppressed photon coupling, while an unsuppressed
portal implies a broader line at higher $m_{\chi_1}$; the argon energy distinguishes
the two. The nuclear sensitivity tests rescale $T_1$ by the rate factor, and
hence the proxy values of $r_\gamma$ by a factor 0.5--1.7. Within the adopted
halo model, daughters around 10 GeV combine a narrow line with much weaker
photon-suppression requirements than sub-GeV daughters. Suppression relative
to the proton-loop reference is an additional model-building requirement;
neutron-dominated couplings alone do not guarantee it.\footnote{A $\phi$
coupling to electrons would open $\chi_1\to\chi_2e^+e^-$ for $\delta>2m_e$ and
tighten the requirements further.}

\emph{An explicit effective-theory benchmark.}\ For H we take $m_\phi=2$ GeV
($q^2/m_\phi^2\simeq0.015$), $g_{12}=0.1$ GeV, $g_p=g_n=1.08\times10^{-5}$,
no electron coupling and $r_\gamma=0.01$, together with $f_{0,i}=1$ and
$\tau_0=200$ Gyr, which requires $g_{011}\simeq8.1\times10^{-19}$ GeV from
$\Gamma_{0\to11}=g_{011}^2(1-4m_{\chi_1}^2/m_{\chi_0}^2)^{1/2}/(32\pi m_{\chi_0})$.
Equation~\eqref{eq:f1} gives $f_0(\tU)=0.933$, $f_1(\tU)=0.0667$ and
$\tau_1=2.0\times10^{15}$ yr, and the resulting $\sigma_n=1.67\times10^{-44}$
cm$^2$ reproduces $\sigma_nf_1$ of Table~\ref{tab:bench}, i.e.\ one
accepted event in 2.84 tonne-years for 50\% acceptance. Because $\tau_1\gg\tU$,
any primordial $\chi_1$ would survive unchanged; the benchmark therefore relies
on a negligible initial abundance or an earlier depletion mechanism. The
ultraviolet portal and its other constraints remain unspecified. In
particular, $r_\gamma=0.01$ ($c_\gamma\simeq1.8\times10^{-10}$ GeV$^{-1}$) is
an independent effective-theory choice; within the scalar portals of
footnote~\ref{fn:rgamma} it requires a cancellation at the $10^{-3}$ level. The emissivity~\eqref{eq:emissivity} lies below the
illustrative line proxy~\eqref{eq:rgamma} for $\tau_\gamma\lesssim3\times10^{27}$ s.

The line lies above the standard nuclear-recoil windows. Whether conventional
xenon limits are evaded must still be checked with the accepted spectrum and
with any additional interaction channels, such as loop-induced elastic
scattering. Extended-window analyses by XENONnT and PandaX-4T
would test it directly. An argon
detector would see the line at 700--810 keV (Fig.~\ref{fig:kin}). Its rate per kg
relative to xenon is uncertain between $\sim0.1$ and $\sim10$, because
$q\simeq230$--245 MeV falls near the first zero of the argon form factor.

\section{Conclusions}
\label{sec:conclusions}
A single event cannot establish a signal, but it can point to the kind of
physics that would produce one. We have argued that the LZ candidate is
naturally explained by exothermic downscattering of a GeV-scale excited state
with an MeV-scale splitting. Rather than relying on a primordial excited
population, we let it be regenerated today by the late decay of a long-lived
parent; a small energy release keeps the daughters bound and the line narrow.

The line position follows from kinematics and is robust. Because the candidate
sits just below the second xenon form-factor zero, the nuclear response
changes the rate by about an order of magnitude but moves the line by at most
a few keV. The price of the scenario is the radiative decay
$\chi_1\to\chi_2\gamma\gamma$. Normalizing to the LZ event fixes its local
emissivity, whatever the production history of $\chi_1$. A rough $\gamma$-ray
estimate then asks for a percent-level photon coupling, relative to the
proton-loop reference, for daughters near 10 GeV, and a far smaller one for
sub-GeV daughters. An unsuppressed coupling survives only for
$m_{\chi_1}\gtrsim40$ GeV, where the line is broader and the argon energy
lower. Our 10 GeV benchmark shows that the required couplings can be realized
in an effective theory. In a vector-portal variant,
the same normalization instead fixes a Galactic positron injection rate at
kinetic energies below 2.4 MeV, to be compared with the 511 keV
emission~\cite{Huh2008}.

Our scenario makes sharp predictions: a line at 700--810 keV in argon and
420--450 keV in germanium, with a ratio to xenon set only by $m_{\chi_1}$. 

\section*{Acknowledgements}
This work was supported by National Research Foundation of Korea (NRF) grants
funded by the Korean government (MSIT), Nos.~RS-2024-00340153 and
RS-2026-25607498.

\end{document}